\documentclass[12pt,a4paper]{article}

\usepackage[a4paper,margin=2.7cm]{geometry}
\usepackage[T1]{fontenc}
\usepackage[utf8]{inputenc}
\usepackage{lmodern}
\usepackage{microtype}
\usepackage{amsmath,amssymb,mathtools,amsthm,bm}
\usepackage{cite}
\usepackage{booktabs,array}
\usepackage{enumitem}
\usepackage[colorlinks=true,linkcolor=blue,citecolor=blue,urlcolor=blue,
  pdftitle={Dissipative non-Abelian fluids from Scherk-Schwarz dimensional reduction},
  pdfauthor={E. Torrente-Lujan},
  pdfkeywords={non-Abelian hydrodynamics, Kaluza-Klein reduction, Scherk-Schwarz compactification, relativistic fluids}]{hyperref}
\usepackage[nameinlink,noabbrev]{cleveref}

\numberwithin{equation}{section}

\newcommand{\dd}{\mathrm{d}}
\newcommand{\ee}{\mathrm{e}}

\newcommand{\heta}{\hat\eta}

\newcommand{\D}{\mathcal{D}}
\newcommand{\Du}{\mathrm{D}_u}

\newcommand{\red}{\mathrm{red}}

\newcommand{\pf}{\mathrm{pf}}

\begin{document}
\title{
\vspace{-2cm}
{\normalsize
\begin{flushright}
UM-FT/26-31415\\
UQBAR-FT/26-2765
\end{flushright}
\vspace{1cm}
}
Dissipative non-Abelian fluids from Scherk--Schwarz dimensional reduction}
\author{E. Torrente-Lujan\\[2mm]
\small FISPAC, Universidad de Murcia, Campus de Espinardo, 30100 Murcia, Spain\\
\small \texttt{etl@um.es}}
\maketitle

\begin{abstract}
We construct a $d$-dimensional dissipative colored fluid by Scherk--Schwarz reduction of a neutral viscous conformal fluid in $D=d+n$ dimensions on an $n$-dimensional unimodular group manifold.  The off-diagonal components of the higher-dimensional stress tensor become non-Abelian color currents, while the higher-dimensional shear tensor induces shear, bulk-like and vector-dissipative structures in the reduced theory.  We derive the map for the equation of state, sound speed, color current, entropy current and first-order transport coefficients.  In particular,
\[
        \eta=\ee^{\alpha\varphi}\cosh\xi\,\heta,\qquad
        \tau=\eta\,\frac{n}{(D-1)(d-1)},\qquad
        \kappa=\eta\sinh^2\xi .
\]
We also spell out the hydrodynamic-frame issue induced by dimensional reduction, discuss the status of the internal rapidity field $\xi$, and give a detailed account of how the second law descends from the parent theory, including the roles of temperature-dependent viscosity, non-unimodular groups and possible choices for $\xi$.  The construction should be regarded as a toy model for non-Abelian dissipative hydrodynamics with the potential of paving the way to direct phenomenological model of, for example, quark--gluon plasma.
\end{abstract}


\section{Introduction}
\label{sec:introduction}

Relativistic hydrodynamics is the long-wavelength effective description of many interacting quantum systems.  Its dynamical variables are local densities, velocities and, when global or gauge symmetries are present, charge currents.  This effective viewpoint is central in the theory of quark--gluon plasma, dense matter, early-universe cosmology and holographic fluid dynamics \cite{Rangamani:2009xk,Hubeny:2011hd,Romatschke:2017ejr,Florkowski:2017olj,Basar:2024}.  When the relevant charges are non-Abelian, the hydrodynamic theory must combine fluid variables with Yang--Mills gauge covariance.  Classical color dynamics and non-Abelian fluid mechanics have been developed from several complementary perspectives \cite{Wong:1970fu,Heinz:1983nx,Heinz:1984yq,Elze:1989un,Jackiw:2000cd,Bistrovic:2002jx,Jackiw:2004nm}.

Dimensional reduction gives a systematic way of generating gauge dynamics from geometry.  The Kaluza--Klein mechanism \cite{Kaluza:1921tu,Klein:1926tv,Overduin:1998pn} and its non-Abelian generalizations \cite{Kerner:1968,DeWitt:1963,Scherk:1979zr,Duff:1986hr,Salam:1981xd,Witten:1981me,Cvetic:2003jy} show how part of the higher-dimensional diffeomorphism group is reorganized as Yang--Mills gauge symmetry in lower dimensions.  This suggests a constructive route to charged hydrodynamics: reduce a higher-dimensional neutral fluid.  Abelian and toroidal reductions of relativistic fluids were studied in \cite{DiDato:2013cla,DiDato:2015dia}.  The closely related non-Abelian construction of \cite{FernandezMelgarejo:2016aet} showed that Kaluza--Klein reduction on a manifold with non-Abelian isometry can produce a colored dissipative fluid coupled to Yang--Mills fields.
The relation to that work is important.  The aim here is to isolate a compact and reusable version of the map in arbitrary $D=d+n$, with explicit attention to the following points: the tangent-space form of the fluid ansatz, the role of unimodularity in current and entropy conservation, the reduced equation of state and sound speed, the hydrodynamic frame inherited after reduction, and the descent of the second law when the parent shear viscosity is allowed to be a local transport function.  We also keep the scalar moduli and the Einstein-frame normalization visible, which makes the result easier to use as a solution-generating map in gravitational reductions.  Recent work on non-Abelian fluid helicities and Lagrangian formulations provides a complementary, non-geometric viewpoint on colored fluids \cite{Nastase:2025hls}.

The purpose of this paper is to give a self-contained formulation of this construction for a general unimodular group manifold and for arbitrary $D=d+n$.  We emphasize three points.  First, the fluid reduction ansatz must be stated in tangent-space components; otherwise the dependence on internal coordinates obscures the lower-dimensional interpretation.  Second, current and entropy conservation require the unimodularity condition $f_{mn}{}^n=0$.  Third, dissipative transport in the reduced theory is highly constrained: a single higher-dimensional shear viscosity induces lower-dimensional shear, bulk and vector-dissipative structures.

We work with mostly-plus Lorentzian signature in the non-compact directions, $u^a u_a=-1$, and with positive internal metric.  Hatted quantities refer to the $D$-dimensional theory.  Greek indices $\mu,\nu$ are $d$-dimensional curved indices, Latin indices $a,b$ are $d$-dimensional tangent indices, $m,n$ label the adjoint of the reduction group, and $\alpha,\beta$ are internal tangent indices.  The dimension split is
\begin{equation}
        D=d+n .
\end{equation}

\section{Geometrical reduction}
\label{sec:geometry}

\subsection{Scherk--Schwarz ansatz}
\label{subsec:ansatz}

Let $G$ be an $n$-dimensional Lie group with left-invariant one-forms $\sigma^m$ satisfying
\begin{equation}
        \dd\sigma^m=-\frac{1}{2}f_{np}{}^m\,\sigma^n\wedge\sigma^p .
        \label{eq:maurer-cartan}
\end{equation}
We use the metric and vielbein ansatz
\begin{align}
        \dd\hat s^2
        &=\ee^{2\alpha\varphi(x)}g_{\mu\nu}(x)\dd x^\mu\dd x^\nu
          +\ee^{2\beta\varphi(x)}h_{mn}(x)\nu^m\nu^n,        \label{eq:metric-ansatz}\\
        \nu^m&=\sigma^m+A^m_\mu(x)\dd x^\mu,\\
        h_{mn}&=\Phi_m{}^\alpha\Phi_n{}^\beta\delta_{\alpha\beta},
        \qquad \det h=1 .                         \label{eq:nu-h}
\end{align}
The constants are chosen so that the lower-dimensional gravitational action is in Einstein frame:
\begin{equation}
        \alpha=-\frac{n}{d-2}\,\beta,
        \qquad
        \alpha^2=\frac{n}{2(D-2)(d-2)},
        \qquad
        \beta^2=\frac{d-2}{2n(D-2)} .               \label{eq:alpha-beta}
\end{equation}
The sign of $\beta$ is fixed by the first relation once $\alpha>0$ is chosen.  We define
\begin{eqnarray}
        F^m&=&\dd A^m+\frac{1}{2}f_{np}{}^m A^n\wedge A^p,\\
        \D X^m&=&\dd X^m+f_{np}{}^m A^n X^p .       \label{eq:F-D}
\end{eqnarray}
Changing the sign convention in \cref{eq:maurer-cartan} is equivalent to $f\mapsto -f$ and does not alter the final physical statements.

The higher-dimensional Einstein equations coupled to a fluid are written in tangent-space form as
\begin{equation}
        \hat R_{AB}-\frac{1}{2}\hat\eta_{AB}\hat R+\hat\Lambda\hat\eta_{AB}
        =\hat T^{\mathrm f}_{AB}.                 \label{eq:D-einstein}
\end{equation}
After reduction they split into lower-dimensional Einstein, Yang--Mills and scalar equations. 

 The part of the $d$-dimensional stress tensor coming from the fluid is simply
\begin{equation}
        T^{\mathrm f}_{ab}=\ee^{2\alpha\varphi}\hat T^{\mathrm f}_{ab}. \label{eq:fluid-stress-reduction}
\end{equation}
The Yang--Mills equation can be written as
\begin{equation}
        \D_b\!\big(\ee^{-2(\alpha-\beta)\varphi}h_{mn}F^{n\,ba}\big)
        = J^{a}_{m,\Phi}+J^{a}_{m,\mathrm f},       \label{eq:ym-eom}
\end{equation}
where $J^{a}_{m,\Phi}$ is the scalar-moduli source and the fluid color current is
\begin{equation}
        J^{a}_{m,\mathrm f}=2\ee^{2\alpha\varphi}\Phi_m{}^\alpha\hat T^{\mathrm f}_{\alpha}{}^{a} . \label{eq:fluid-color-current-def}
\end{equation}
The factor of two is a normalization convention inherited from the mixed Einstein equation.  With this convention the gauge field couples directly to the mixed components of the higher-dimensional fluid stress tensor.

\subsection{Unimodularity and reduced conservation laws}
\label{subsec:unimodular}

For a vector $\hat X^A=(\hat X^a,\hat X^\alpha)$ whose tangent-space components depend only on $x^\mu$, the higher-dimensional divergence reduces to
\begin{equation}
\hat\nabla_A\hat X^A
=\ee^{-2\alpha\varphi}\nabla_\mu\!\big(\ee^{\alpha\varphi}e_a{}^\mu\hat X^a\big)
+\ee^{-\beta\varphi}\Phi_\alpha{}^m\,f_{mn}{}^n\hat X^\alpha .
\label{eq:divergence-reduction}
\end{equation}
Consequently, for unimodular groups,
\begin{equation}
        f_{mn}{}^n=0,                              \label{eq:unimodular}
\end{equation}
any conserved higher-dimensional vector induces a conserved lower-dimensional vector
\begin{equation}
        X^\mu_{\red}=\ee^{\alpha\varphi}e_a{}^\mu\hat X^a,
        \qquad
        \hat\nabla_A\hat X^A=0\quad\Rightarrow\quad \nabla_\mu X^\mu_{\red}=0 .
        \label{eq:reduced-current}
\end{equation}
This observation is used below for both charge and entropy currents.  If \cref{eq:unimodular} is not imposed, the lower-dimensional current conservation laws acquire an anomalous source proportional to $f_{mn}{}^n$.

\section{Fluid reduction ansatz}
\label{sec:fluid-ansatz}

The higher-dimensional velocity can be  decomposed in tangent-space components as
\cite{FernandezMelgarejo:2016aet}
\begin{equation}
        \hat u^a=u^a\cosh\xi,
        \qquad
        \hat u^\alpha=n^\alpha\sinh\xi,              \label{eq:velocity-ansatz}
\end{equation}
with
\begin{equation}
        u^a u_a=-1,
        \qquad
        n^\alpha n_\alpha=1 .                       \label{eq:normalizations}
\end{equation}
Thus $\hat u^A\hat u_A=-1$.  The rapidity-like field $\xi(x)$ measures the amount of higher-dimensional fluid velocity lying in the internal directions, while $n^\alpha(x)$ determines its color orientation.  

We also define
\begin{equation}
        n_m=\Phi_m{}^\alpha n_\alpha,
        \qquad
        h^{mn}n_m n_n=1.                             \label{eq:nm}
\end{equation}
All scalar thermodynamic fields are assumed to depend only on $x^\mu$.  In curved components, the velocity contains the gauge connection through the metric ansatz; this is the origin of non-Abelian charge in the reduced theory.

The status of $\xi$ deserves a separate comment.  In the minimal version of the construction, $\xi$ is a constant label of the reduction sector and measures the internal boost of the parent fluid.  More generally it may be treated as a prescribed slowly varying background, in which case its gradients appear as fixed sources in the first-order transport map.  Promoting $\xi(x)$ to an independent hydrodynamic field is possible only after supplying an additional equation of motion or constitutive relation; otherwise the reduced hydrodynamic system is underdetermined.  Unless stated otherwise, thermodynamic derivatives such as the sound speed below are taken at fixed $\xi$ and fixed scalar moduli.

A useful kinematic identity follows from \cref{eq:divergence-reduction}.  The higher-dimensional expansion is
\begin{equation}
        \hat\theta\equiv\hat\nabla_A\hat u^A
        =\ee^{-\alpha\varphi}\cosh\xi
        \left[\theta+\Du\log\big(\ee^{\alpha\varphi}\cosh\xi\big)\right],
        \qquad
        \theta=\nabla_\mu u^\mu,
        \quad \Du\equiv u^\mu\nabla_\mu .                    \label{eq:theta-reduction}
\end{equation}
This non-homogeneous relation is the basic reason why a simple higher-dimensional dissipative term generates several lower-dimensional first-order structures.
%

For reference, the central entries of the hydrodynamic dictionary are summarized in \cref{tab:dictionary}.  The table separates three layers of the construction: the geometrical fields produced by the Scherk--Schwarz ansatz, the perfect-fluid thermodynamic variables, and the first-order dissipative data.
\begin{table}[ht]
\centering
\small
\renewcommand{\arraystretch}{1.2}
\begin{tabular}{>{\raggedright\arraybackslash}p{0.23\linewidth} >{\raggedright\arraybackslash}p{0.35\linewidth} >{\raggedright\arraybackslash}p{0.32\linewidth}}
\toprule
Higher-dimensional object & Reduced object & Interpretation \\
\midrule
$\hat g_{MN}$ & $g_{\mu\nu}$, $A^m_\mu$, $\varphi$, $h_{mn}$ & gravity, Yang--Mills fields, volume scalar and internal moduli \\
$\hat u^A$ & $u^a$, $n^\alpha$, $\xi$ & external velocity, color orientation and internal rapidity \\
$\hat T_{ab}^{\mathrm f}$ & $T^{\mathrm f}_{ab}=\ee^{2\alpha\varphi}\hat T^{\mathrm f}_{ab}$ & reduced fluid stress tensor \\
$\hat T_{a\alpha}^{\mathrm f}$ & $J^a_{m,\mathrm f}=2\ee^{2\alpha\varphi}\Phi_m{}^\alpha\hat T_{\alpha}{}^a$ & non-Abelian fluid current \\
$\hat p$, $\hat\rho$ & $p=\ee^{2\alpha\varphi}\hat p$, $\rho=\ee^{2\alpha\varphi}(\hat\rho\cosh^2\xi+\hat p\sinh^2\xi)$ & reduced pressure and energy density at fixed $\xi$, $\varphi$ \\
$\hat\eta$ & $\eta$, $\tau$, $\kappa$ & induced shear, bulk-like and vector transport coefficients \\
$\hat s^A$ & $s^\mu=\ee^{\alpha\varphi}e_a{}^\mu\hat s^a$ & entropy current inherited from the parent theory \\
\bottomrule
\end{tabular}
\caption{Core entries of the reduction dictionary.  The explicit transport coefficients are given in \cref{eq:eta-tau,eq:kappa}.}
\label{tab:dictionary}
\end{table}

\section{Perfect parent fluid}
\label{sec:perfect}

Consider first a neutral perfect fluid in $D$ dimensions,
\begin{equation}
        \hat T^{\pf}_{AB}=\hat w\,\hat u_A\hat u_B+\hat p\,\hat\eta_{AB},
        \qquad
        \hat w=\hat\rho+\hat p.                       \label{eq:perfect-parent}
\end{equation}
The reduced stress tensor keeps the perfect-fluid form
\begin{equation}
        T^{\pf}_{\mu\nu}=w u_\mu u_\nu+p g_{\mu\nu},  \label{eq:perfect-lower}
\end{equation}
where
\begin{align}
        w&=\ee^{2\alpha\varphi}\hat w\cosh^2\xi,      \label{eq:w-map}\\
        p&=\ee^{2\alpha\varphi}\hat p,                \label{eq:p-map}\\
        \rho&=\ee^{2\alpha\varphi}\left(\hat\rho\cosh^2\xi+
        \hat p\sinh^2\xi\right).                      \label{eq:rho-map}
\end{align}
If $\xi$ and $\varphi$ are held fixed in the local thermodynamic variation, the sound speeds obey
\begin{equation}
        c_s^2\equiv\frac{\partial p}{\partial\rho}
        =\frac{1}{\cosh^2\xi\,\hat c_s^{-2}+\sinh^2\xi} .
        \label{eq:sound-speed}
\end{equation}
Equation~\eqref{eq:sound-speed} is a partial derivative on the reduced equilibrium surface at fixed $\xi$ and fixed scalar moduli.  If $\xi$ or $\varphi$ are promoted to thermodynamic variables, the denominator receives additional susceptibility terms.  For example, a variation $\dd\xi\ne0$ produces contributions proportional to $\partial_\xi p$ and $\partial_\xi\rho$, and the result is no longer determined by the parent sound speed alone.  This is why the present work treats $\xi$ as a reduction label or external background unless an extra constitutive equation is specified.

For a conformal parent fluid, $\hat\rho=(D-1)\hat p$ and $\hat c_s^2=1/(D-1)$, this gives
\begin{equation}
        \frac{p}{\rho}=\frac{1}{D\cosh^2\xi-1}
        =\frac{1}{(d-1)+n\cosh^2\xi+d\sinh^2\xi}.       \label{eq:reduced-conformal-eos}
\end{equation}
Thus the reduced fluid is generically non-conformal even when the parent fluid is conformal.

The perfect-fluid contribution to the non-Abelian current is obtained from \cref{eq:fluid-color-current-def}:
\begin{equation}
        J^{\mu}_{m,\pf}=q_0 n_m u^\mu,
        \qquad
        q_0=2\ee^{2\alpha\varphi}\hat w\sinh\xi\cosh\xi
            =2w\tanh\xi .                              \label{eq:perfect-current}
\end{equation}
The charge density measured by comoving observers is $Q_m=q_0 n_m$.  In a gauge-covariant notation the non-Abelian continuity equation is
\begin{equation}
        \D_\mu J^\mu_m=0,                               \label{eq:cov-current}
\end{equation}
where the precise covariant derivative depends on whether the adjoint index is placed upstairs or downstairs.

\subsection{Entropy current and thermodynamic Wong equation}
\label{subsec:entropy-perfect}

Assume that the parent perfect fluid admits a local-equilibrium entropy current $\hat s^A=\hat s\hat u^A$ satisfying $\hat\nabla_A\hat s^A=0$ and
\begin{equation}
        \hat w=\hat T\hat s,
        \qquad
        \dd\hat p=\hat s\,\dd\hat T .                  \label{eq:parent-thermo}
\end{equation}
By \cref{eq:reduced-current}, the current
\begin{equation}
        s^\mu_{0}=\ee^{\alpha\varphi}\hat s\cosh\xi\,u^\mu              \label{eq:reduced-entropy-raw}
\end{equation}
is conserved.  In the charged lower-dimensional theory it is useful to introduce color chemical potentials $\mu^m$ and write the thermodynamic relations as
\begin{equation}
        \rho+p=Ts+\mu^mQ_m,
        \qquad
        \dd p=s\,\dd T+Q_m\,\dd\mu^m .                 \label{eq:charged-thermo}
\end{equation}
The canonical perfect-fluid entropy current is then
\begin{equation}
        s^\mu=\frac{p u^\mu-T^{\mu\nu}_{\pf}u_\nu-\mu^mJ^\mu_{m,\pf}}{T}
        =s u^\mu .                                      \label{eq:canonical-entropy}
\end{equation}
Using the equations of motion, entropy conservation can be written as the thermodynamic Wong relation
\begin{equation}
        \mu^m\left(\Du Q_m+Q_m\theta\right)=u_\nu\nabla_\mu T^{\mu\nu}_{\pf} .
        \label{eq:wong-thermo}
\end{equation}
In the absence of external force density the right-hand side vanishes and the color charge is convected with the fluid.

\section{First-order conformal viscous parent fluid}
\label{sec:viscous}

We now take the parent fluid to be conformal and first-order viscous,
\begin{equation}
        \hat T^{\mathrm f}_{AB}=\hat p\,\hat\eta_{AB}+\hat w\,\hat u_A\hat u_B
        -\hat\eta\,\hat\sigma_{AB},                    \label{eq:conformal-parent}
\end{equation}
with
\begin{equation}
        \hat\rho=(D-1)\hat p,\qquad \hat w=D\hat p,
        \qquad
        \hat T^A{}_A=0 .                                \label{eq:conformal-trace}
\end{equation}
The shear tensor is
\begin{equation}
        \hat\sigma_{AB}
        =\hat\Pi_A{}^C\hat\Pi_B{}^E
        \left(\hat\nabla_{(C}\hat u_{E)}-\frac{1}{D-1}\hat\Pi_{CE}\hat\theta\right),
        \qquad
        \hat\Pi_{AB}=\hat\eta_{AB}+\hat u_A\hat u_B .   \label{eq:shear-parent}
\end{equation}
The reduced stress tensor is
\begin{equation}
        T^{\mathrm f}_{ab}=T^{\pf}_{ab}-\ee^{2\alpha\varphi}\hat\eta\,\hat\sigma_{ab}. \label{eq:viscous-reduction-def}
\end{equation}
The part of $\hat\sigma_{ab}$ proportional to the lower-dimensional shear and expansion is universal:
\begin{equation}
        \hat\sigma_{ab}
        =\ee^{-\alpha\varphi}\cosh\xi\,\sigma_{ab}
        +\ee^{-\alpha\varphi}\cosh\xi
        \left(\frac{1}{d-1}-\frac{1}{D-1}\right)\Pi_{ab}\theta
        +\mathcal{S}_{ab},                              \label{eq:universal-shear-map}
\end{equation}
where
\begin{equation}
        \Pi_{ab}=\eta_{ab}+u_a u_b,
        \qquad
        \sigma_{ab}=\Pi_a{}^c\Pi_b{}^e
        \left(\nabla_{(c}u_{e)}-\frac{1}{d-1}\Pi_{ce}\theta\right). \label{eq:lower-shear}
\end{equation}
The tensor $\mathcal{S}_{ab}$ contains the remaining first-order terms generated by the scalar moduli, gauge fields, gradients of $\xi$ and acceleration; it is displayed compactly in \cref{app:derivative-map}.  Equation \eqref{eq:universal-shear-map} immediately gives the transport coefficients
\begin{eqnarray}
        \eta&=&\ee^{\alpha\varphi}\cosh\xi\,\hat\eta,\\
        \tau&=&\eta\,\frac{n}{(D-1)(d-1)}.         \label{eq:eta-tau}
\end{eqnarray}
Here $\eta$ is the lower-dimensional shear viscosity and $\tau$ is the coefficient multiplying the induced bulk-like term $-\tau\theta\Pi_{ab}$.  Since a conformal parent fluid has no bulk viscosity, this term is entirely geometrical: it is produced by reducing from $D$ to $d$ dimensions.

It is useful to decompose the full reduced dissipative correction as
\begin{equation}
        \pi_{ab}
        \equiv T^{\mathrm f}_{ab}-T^{\pf}_{ab}
        =-\eta\sigma_{ab}-\tau\theta\Pi_{ab}
          +q_{(a}u_{b)}+\delta p_{\rm geo}\Pi_{ab}+\delta\rho_{\rm geo}u_a u_b
          +\pi^{\rm geo}_{ab}.                         \label{eq:pi-decomposition}
\end{equation}

The quantities with the subscript ``geo'' are determined by $\mathcal{S}_{ab}$ and are not independent phenomenological coefficients.  The transverse vector term takes the form
\begin{eqnarray}
        q_a&=&\kappa\,\Pi_a{}^b\mathcal{V}_b,\\
        \kappa&=&\eta\sinh^2\xi,                  \label{eq:kappa}
\end{eqnarray}
where $\mathcal{V}_b$ is a fixed covariant combination of the acceleration, color-electric field $E^m_b=F^m_{bc}u^c$, scalar gradients and $\Du n_m$.  Therefore the lower-dimensional first-order data are not arbitrary: they are locked by the dimensional reduction map.

The color current also receives a dissipative term,
\begin{equation}
        J^\mu_m=J^\mu_{m,\pf}-2\ee^{2\alpha\varphi}\hat\eta\,e_a{}^\mu\Phi_m{}^\alpha\hat\sigma_{\alpha}{}^{a}. \label{eq:viscous-current}
\end{equation}
Writing
\begin{equation}
        J^\mu_m=Q_m u^\mu+\nu^\mu_m,
    \end{equation}
  with    
  \begin{equation} 
        u_\mu\nu^\mu_m=0,                              \label{eq:current-decomp}
\end{equation}
one finds
\begin{equation}
        Q_m=q_0 n_m+\eta Q^{(1)}_m,
           \end{equation}
  with    
  \begin{equation} 
        \nu^\mu_m=\eta\nu^{(1)\mu}_m,                   \label{eq:diss-current-components}
\end{equation}
with $Q^{(1)}_m$ and $\nu^{(1)\mu}_m$ fixed by \cref{eq:velocity-ansatz,eq:shear-parent}.  No new color-diffusion transport coefficient is introduced independently of $\hat\eta$.

\subsection{Compact statement of the transport map}
\label{subsec:proposition}

The dissipative part of the construction can be summarized as follows.
(First-order Scherk--Schwarz transport map)
Let a neutral conformal parent fluid in $D=d+n$ dimensions have first-order stress tensor \eqref{eq:conformal-parent}, shear tensor \eqref{eq:shear-parent} and non-negative shear viscosity $\hat\eta$.  Let the reduction be performed on a unimodular group manifold, with the velocity ansatz \eqref{eq:velocity-ansatz}.  Then the $d$-dimensional colored fluid has a non-Abelian current \eqref{eq:fluid-color-current-def}, equation of state \eqref{eq:p-map}--\eqref{eq:rho-map}, and first-order dissipative stress tensor of the form
\begin{equation}
        \pi_{ab}=-\eta\sigma_{ab}-\tau\theta\Pi_{ab}+q_{(a}u_{b)}
        +\delta p_{\rm geo}\Pi_{ab}+\delta\rho_{\rm geo}u_a u_b+\pi^{\rm geo}_{ab},
\end{equation}
where the independent coefficients inherited from the parent viscosity are
\begin{equation}
        \eta=\ee^{\alpha\varphi}\cosh\xi\,\hat\eta,
        \qquad
        \tau=\eta\frac{n}{(D-1)(d-1)},
        \qquad
        \kappa=\eta\sinh^2\xi .
\end{equation}
All remaining first-order terms are fixed geometrically by $F^m_{\mu\nu}$, $h_{mn}$, $\varphi$, $n^\alpha$ and possible gradients of $\xi$.

This is the main practical output of the paper.  It should be read as a constrained embedding of non-Abelian first-order hydrodynamics into a higher-dimensional neutral system, possibly not as the most general colored-fluid constitutive relation allowed by symmetry.

\subsection{Hydrodynamic frame after dimensional reduction}
\label{subsec:frame}

The hydrodynamic frame is not generally preserved by the reduction.  Suppose that the parent stress tensor is in a Landau-type frame,
\begin{equation}
        \hat T^{\mathrm f}_{AB}\hat u^B=-\hat\rho_{\rm L}\hat u_A .        \label{eq:parent-landau}
\end{equation}
Using \cref{eq:velocity-ansatz,eq:fluid-stress-reduction,eq:fluid-color-current-def}, the external component implies
\begin{equation}
        T^{\mathrm f}_{ab}u^b
        =-\ee^{2\alpha\varphi}\hat\rho_{\rm L}u_a
        -\frac{1}{2}\tanh\xi\,J^{\mathrm f}_{a m}n^m .                 \label{eq:landau-not-preserved}
\end{equation}
Therefore $u^a$ is an eigenvector of the reduced stress tensor only when $\xi=0$, or when the color-current projection $J^{\mathrm f}_{a m}n^m$ is purely longitudinal, or when this projection vanishes.  The perfect fluid satisfies this automatically because $J^\mu_{m,\pf}\propto u^\mu$, but the viscous current in \cref{eq:viscous-current} typically has a transverse component. 

 Consequently the reduced theory is naturally written in a general hydrodynamic frame rather than forced into the Landau frame.  This is consistent with modern analyses showing that first-order relativistic hydrodynamics can be stable in suitable non-Landau frames, while the Landau--Lifshitz frame itself is not in the stable class considered in \cite{Kovtun:2019hdm}.

\subsection{Entropy production}
\label{subsec:entropy-viscous}

For the conformal viscous parent fluid we take the canonical first-order entropy current
\begin{equation}
        \hat s^A=\frac{\hat p\,\hat u^A-\hat T^{AB}\hat u_B}{\hat T}.      \label{eq:parent-entropy-viscous}
\end{equation}
In the Landau frame of the parent theory, $\hat\sigma_{AB}\hat u^B=0$, this reduces to $\hat s\hat u^A$ at ideal order, but writing it as \eqref{eq:parent-entropy-viscous} is useful because the cancellation with the viscous stress tensor is explicit.  Using stress-tensor conservation, $\hat\nabla_A\hat T^{AB}=0$, and the local thermodynamic relations of the parent fluid, one obtains at the order controlled by first-order hydrodynamics
\begin{equation}
        \hat\nabla_A\hat s^A
        =-\frac{1}{\hat T}\hat\pi^{AB}\hat\nabla_A\hat u_B
        =\frac{\hat\eta}{\hat T}\,\hat\sigma_{AB}\hat\sigma^{AB}
        \ge 0,\qquad \hat\pi_{AB}=-\hat\eta\hat\sigma_{AB}.             \label{eq:parent-entropy-prod}
\end{equation}
Thus the parent second law requires $\hat\eta\ge0$.  This statement does not require $\hat\eta$ to be a spacetime constant.  For a conformal parent fluid one typically has $\hat\eta=\hat\eta_0\hat T^{D-1}$; the gradient of $\hat\eta(\hat T)$ would contribute only beyond the first-order entropy-production analysis once the constitutive relation is consistently truncated.  Therefore the required local condition is simply $\hat\eta(\hat T)\ge0$.

For a unimodular group, \cref{eq:divergence-reduction} gives the reduced entropy current
\begin{equation}
        s^\mu_{\rm inh}=\ee^{\alpha\varphi}e_a{}^\mu\hat s^a,       \label{eq:inherited-entropy-current}
\end{equation}
and the entropy production descends as
\begin{equation}
        \nabla_\mu s^\mu_{\rm inh}
        =\ee^{2\alpha\varphi}\frac{\hat\eta}{\hat T}\,
        \hat\sigma_{AB}\hat\sigma^{AB}\ge0 .             \label{eq:reduced-entropy-prod}
\end{equation}
This is the strongest form of the result: positivity is inherited from the positive quadratic form in the parent theory and from the unimodular reduction of a vector divergence.

The same statement can be rewritten in lower-dimensional charged variables.  With
\begin{equation}
        J^\mu_m=Q_m u^\mu+\nu^\mu_m,\qquad u_\mu\nu^\mu_m=0,
\end{equation}
and with the charged Gibbs--Duhem relations
\begin{equation}
        \rho+p=Ts+\mu^mQ_m,
        \qquad
        \dd p=s\,\dd T+Q_m\,\dd\mu^m,
\end{equation}
the canonical charged entropy current can be written schematically as
\begin{equation}
        s^\mu_{\rm can}
        =\frac{p u^\mu-T^{\mu\nu}u_\nu-\mu^mJ^\mu_m}{T}
        =s u^\mu-\frac{\mu^m}{T}\nu^\mu_m+\hbox{frame terms}.          \label{eq:charged-canonical-entropy}
\end{equation}
The phrase ``frame terms'' is important: because \cref{eq:landau-not-preserved} shows that the reduced variables are not generally in Landau frame, the canonical entropy current and the inherited entropy current agree only after the appropriate frame choice is made.  Their divergences nevertheless encode the same inequality, which may be written as
\begin{equation}
        \mu^m\left(\Du Q_m+Q_m\theta\right)-u_\nu\nabla_\mu T^{\mu\nu}_{\pf}
        \ge 0,                                           \label{eq:diss-wong-ineq}
\end{equation}
with equality in the perfect-fluid limit.  In the absence of external force density, this reduces to a dissipative version of the thermodynamic Wong equation.

It is also useful to state what happens outside the assumptions used above.
\begin{enumerate}[label=(\roman*),leftmargin=2.2em]
\item \emph{Constant $\xi$.}  This is the cleanest sector.  The reduced equation of state, sound speed and transport coefficients are algebraic functions of $\xi$, and \cref{eq:reduced-entropy-prod} is a direct non-negative entropy-production law.
\item \emph{Background $\xi(x)$.}  The gradients $\nabla_\mu\xi$ enter the geometrical pieces $\delta p_{\rm geo}$, $\delta\rho_{\rm geo}$, $q_a$ and $\pi^{\rm geo}_{ab}$.  Positivity still follows from the uplifted expression \eqref{eq:reduced-entropy-prod}, provided $\xi(x)$ is treated as prescribed data of a consistent higher-dimensional configuration.  From the purely lower-dimensional viewpoint these gradients are external sources.
\item \emph{Dynamical $\xi(x)$.}  If $\xi$ is promoted to a hydrodynamic field, the second law cannot close with the variables used here alone.  One must add either an evolution equation for $\xi$, a Josephson-like relation, or an enlarged thermodynamic first law with the susceptibility conjugate to $\xi$.
\item \emph{Non-unimodular groups.}  If $f_{mn}{}^n\ne0$, the divergence formula gives
\begin{equation}
        \nabla_\mu s^\mu_{\rm inh}
        =\ee^{2\alpha\varphi}\hat\nabla_A\hat s^A
        -\ee^{(2\alpha-\beta)\varphi}\Phi_\alpha{}^m f_{mn}{}^n\hat s^\alpha .      \label{eq:non-unimodular-entropy}
\end{equation}
The last term has no definite sign in general.  Hence unimodularity is not a cosmetic assumption; it is what makes the inherited second law automatic.
\item \emph{Second-order or causal completions.}  The present construction is first order.  A causal Israel--Stewart-type uplift would require reducing the parent relaxation variables and their entropy-current corrections.  Such an extension may preserve the geometric locking of transport coefficients, but it is a distinct problem.
\end{enumerate}

Hence the reduced second law is best understood as an inherited theorem in the unimodular, upliftable sector, and as a set of lower-dimensional constraints once one treats the reduced variables independently of their higher-dimensional origin.

\section{Example: $d=4$ and $G=SU(2)$}
\label{sec:example-su2}

A simple concrete case is obtained by choosing
\begin{equation}
        d=4,
        \qquad
        G=SU(2),
        \qquad
        n=3,
        \qquad
        D=7 .
\end{equation}
For an isotropic internal metric $h_{mn}=\delta_{mn}$, constant scalar $\varphi$ and constant internal rapidity $\xi$, the three structure constants may be written as $f_{mn}{}^p=\epsilon_{mn}{}^p$, and unimodularity is automatic.  The reduced conformal-parent equation of state becomes
\begin{equation}
        \frac{p}{\rho}=\frac{1}{7\cosh^2\xi-1}
        =\frac{1}{3+3\cosh^2\xi+4\sinh^2\xi}.          \label{eq:su2-eos}
\end{equation}
For $\xi=0$ this gives $p/\rho=1/6$, as expected from a seven-dimensional conformal parent fluid viewed without internal motion.  Non-zero $\xi$ lowers this ratio further and therefore increases the non-conformality of the effective four-dimensional fluid.

The current and transport coefficients reduce to
\begin{equation}
        J^\mu_{m,\pf}=q_0 n_m u^\mu,
        \qquad
        q_0=2\ee^{2\alpha\varphi}\hat w\sinh\xi\cosh\xi,
\end{equation}
with
\begin{equation}
        \eta=\ee^{\alpha\varphi}\cosh\xi\,\hat\eta,
        \qquad
        \tau=\eta\frac{3}{(7-1)(4-1)}=\frac{\eta}{6},
        \qquad
        \kappa=\eta\sinh^2\xi .                         \label{eq:su2-transport}
\end{equation}
Thus even this minimal $SU(2)$ example produces a colored fluid with shear, a geometrically induced bulk-like term and a vector-dissipative sector, all controlled by the single parent coefficient $\hat\eta$ and by the internal rapidity $\xi$. 

 This example is useful for comparison with formulations that use $SU(2)$ as the simplest non-Abelian gauge group, while the derivation above applies to any unimodular group.

\section{Discussion and conclusions}
\label{sec:conclusion}

Scherk--Schwarz reduction on a unimodular group manifold maps a neutral dissipative fluid in $D=d+n$ dimensions to a lower-dimensional fluid carrying non-Abelian charge.  The fluid color current is the mixed internal/external component of the higher-dimensional stress tensor, and its conservation is a consequence of the higher-dimensional Bianchi identity together with the unimodularity condition.  The same condition ensures that a non-negative entropy production law in the parent theory descends to a non-negative entropy production law in the reduced theory.

Among the main results, it is the constrained transport map summarized in \cref{subsec:proposition}.  For a perfect parent fluid, the reduced fluid remains perfect but acquires non-Abelian charge and a modified equation of state.  A conformal parent fluid is not generally conformal after reduction; the internal kinetic component parametrized by $\xi$ changes both the equation of state and the sound speed.  For a first-order conformal viscous parent fluid, the reduced transport sector is more structured: shear viscosity, an induced bulk-like coefficient and a vector-dissipative coefficient are fixed by a single higher-dimensional shear viscosity according to \cref{eq:eta-tau,eq:kappa}.  The lower-dimensional theory therefore contains fewer independent transport coefficients than the most general non-Abelian dissipative fluid compatible with symmetries.

The construction also clarifies a frame issue.  A Landau-frame parent fluid does not generically reduce to a Landau-frame lower-dimensional fluid, because the projected color current in \cref{eq:landau-not-preserved} contributes to $T_{ab}u^b$.  This is not a pathology; it means that the natural reduced variables are those supplied by the higher-dimensional geometry.  Forcing the result into a preferred lower-dimensional frame would reshuffle the transport data and obscure the geometric origin of the coefficients.

The model is better viewed as a controlled formal laboratory for colored dissipative fluids, scalar moduli and gauge/fluid couplings.  Its value is that any higher-dimensional solution of the neutral Einstein-fluid system satisfying the ansatz gives a lower-dimensional solution with non-Abelian hydrodynamic structure.

The first-order nature of the construction deserves one final caveat.  Questions of causality and stability require either a suitable first-order frame or a second-order completion.  General-frame first-order hydrodynamics can possess stable sectors \cite{Kovtun:2019hdm}, but the reduction of a full Israel--Stewart theory would introduce additional relaxation variables and entropy-current terms \cite{Israel:1979wp}.  

Natural extensions therefore include second-order causal hydrodynamics, anomalous transport, reductions of string-effective actions in which scalar truncations are consistent, and explicit cosmological or black-brane solutions.  

These extensions would clarify which parts of a general non-Abelian hydrodynamic gradient expansion can be geometrized by dimensional oxidation.

\appendix

\section{Useful reduction formulae}
\label{app:geometry-formulae}

In the conventions of \cref{sec:geometry}, the determinant factorizes as
\begin{equation}
        \sqrt{-\hat g}=\sqrt{-g}\,\sqrt{g_G}\,\ee^{2\alpha\varphi},      \label{eq:det-factor}
\end{equation}
where $g_G$ is the determinant of the internal group-frame metric.  The scalar kinetic tensor is
\begin{equation}
        P_{a\alpha\beta}=\frac{1}{2}\left(\Phi_\alpha{}^m\D_a\Phi_{m\beta}
        +\Phi_\beta{}^m\D_a\Phi_{m\alpha}\right),
        \qquad
        P_a{}^\alpha{}_{\alpha}=0,                       \label{eq:P-def}
\end{equation}
and the antisymmetric composite connection is
\begin{equation}
        Q_{a\alpha\beta}=\frac{1}{2}\left(\Phi_\alpha{}^m\D_a\Phi_{m\beta}
        -\Phi_\beta{}^m\D_a\Phi_{m\alpha}\right).          \label{eq:Q-def}
\end{equation}
The purely bosonic lower-dimensional stress tensor takes the schematic form
\begin{align}
        T^{\rm KK}_{ab}
        &=\partial_a\varphi\partial_b\varphi-\frac{1}{2}\eta_{ab}(\partial\varphi)^2
        +P_{a\alpha\beta}P_b{}^{\alpha\beta}
        -\frac{1}{2}\eta_{ab}P_{c\alpha\beta}P^{c\alpha\beta}
        \nonumber\\
        &\quad
        +\ee^{-2(\alpha-\beta)\varphi}h_{mn}
        \left(F^m{}_{ac}F^{n}{}_{b}{}^{c}-\frac{1}{4}\eta_{ab}F^m{}_{cd}F^{ncd}\right)
        -\eta_{ab}V(\varphi,h).                         \label{eq:KK-stress}
\end{align}
The scalar potential is a quadratic expression in the structure constants and the internal metric; its detailed sign depends on the convention chosen for \cref{eq:maurer-cartan}.  It vanishes for toroidal reductions.

\section{First-order derivative map}
\label{app:derivative-map}

The reduced shear map can be written compactly as
\begin{equation}
        \hat\sigma_{ab}=\ee^{-\alpha\varphi}\cosh\xi\,\sigma_{ab}
        +\ee^{-\alpha\varphi}\cosh\xi
        \left(\frac{1}{d-1}-\frac{1}{D-1}\right)\Pi_{ab}\theta
        +\mathcal{S}_{ab}.                              \label{eq:sigma-map-app}
\end{equation}
The remaining term is
\begin{align}
\mathcal{S}_{ab}
&= \frac{1}{D-1}\Pi_{ab}\ee^{-\alpha\varphi}\cosh\xi\,
        \Du\log\big(\ee^{\alpha\varphi}\cosh\xi\big)
        -\sinh^2\xi\,u_a u_b\,\hat\theta
        \nonumber\\
&\quad
        -\ee^{-\alpha\varphi}\sinh^2\xi\cosh\xi\,u_{(a}a_{b)}
        -u_{(a}p_{b)},                                  \label{eq:S-ab}
\end{align}
where
\begin{equation}
        a_b=\Du u_b,
        \qquad
        \hat\theta=\ee^{-\alpha\varphi}\cosh\xi
        \left[\theta+\Du\log\big(\ee^{\alpha\varphi}\cosh\xi\big)\right],  \label{eq:theta-app}
\end{equation}
and
\begin{align}
        p_b
        &= -\sinh^2\xi\cosh\xi\,\ee^{-\alpha\varphi}u_b
           \left(\beta\Du\varphi+P_{u}\right)
        +\frac{1}{2}\sinh\xi\,\ee^{-(2\alpha-\beta)\varphi}E^m_b n_m
        \nonumber\\
        &\quad
        -\sinh^2\xi\cosh\xi\,\ee^{-\alpha\varphi}\Pi_b{}^c
          \left(\beta\nabla_c\varphi+P_c\right).         \label{eq:p-b}
\end{align}
Here
\begin{equation}
        E^m_b=F^m_{bc}u^c,
        \qquad
        P_c=P_{c\alpha\beta}n^\alpha n^\beta,
        \qquad
        P_u=u^cP_{c\alpha\beta}n^\alpha n^\beta .        \label{eq:defs-pb}
\end{equation}
Equations \eqref{eq:sigma-map-app}--\eqref{eq:defs-pb} are sufficient to reconstruct the full first-order reduced stress tensor from \cref{eq:viscous-reduction-def}.  They also show explicitly that the gauge field enters the vector dissipative sector through the color-electric projection $E^m_b$.

\section*{Acknowledgements}
The work of ETL has been supported in part by the Ministerio de Educaci\'on y Ciencia, grants FIS2025-24924,
 Universidad de Murcia project E024-018 and Fundacion Seneca (21257/PI/24). 
 The author thanks the Harvard University Physics Department for hospitality during a early part of the development of this work. Fruitful conversations with J.J. Fernandez-Melgarejo are grateful acknowledged

\end{document}